\documentclass[usenatbib]{mn2e}
\usepackage{graphicx,color,epsfig}

\newcommand{\be}{\begin{equation}}
\newcommand{\ee}{\end{equation}}

\newcommand{\apj}{ApJ}

\newcommand{\mnras}{MNRAS}
\newcommand{\aap}{A\&A}
\newcommand{\araa}{ARA\&A}
\newcommand{\apjl}{ApJL}
\newcommand{\aj}{AJ}
\newcommand{\nat}{Nature}

\def\ltsima{$\; \buildrel < \over \sim \;$}
\def\simlt{\lower.5ex\hbox{\ltsima}}
\def\gtsima{$\; \buildrel > \over \sim \;$}
\def\simgt{\lower.5ex\hbox{\gtsima}}
\def\sgra{Sgr~A$^*$}

\def\msun{{\,{\rm M}_\odot}}

\def\del#1{{}}

\title[Feedback outflows and satellite galaxies]{A link between feedback
  outflows and satellite galaxy suppression}

\author[S. Nayakshin, M. I. Wilkinson]{Sergei Nayakshin and Mark I. Wilkinson\\ 
Department of Physics \& Astronomy,
  University of Leicester, Leicester, LE1 7RH, UK\\
{E-mail:~} {\rm Sergei.Nayakshin@le.ac.uk}}

\begin{document}

\date{Received}

\pagerange{\pageref{firstpage}--\pageref{lastpage}} \pubyear{2008}

\maketitle

\label{firstpage}

\begin{abstract}
We suggest a direct, causal link between the two "missing" baryon problems of
contemporary galaxy formation theory: (1) that large galaxies (such as the
Milky Way) are known to contain too little gas and stars and (2) that too few
dwarf satellite galaxies are observed around large galaxies compared with
cosmological simulations. The former can be explained by invoking some
energetic process -- most likely AGN or star formation feedback -- which
expels to infinity a significant fraction of the gas initially present in the
proto-galaxy, while the latter problem is usually explained by star formation
feedback inside the dwarf or tidal and ram pressure stripping of the gas from
the satellite galaxy by its parent. Here we point out that the host galaxy
``missing'' baryons, if indeed ejected at velocities of hundreds to a thousand
km s$^{-1}$, must also affect smaller satellite galaxies by stripping or
shocking the gas there. We estimate the fraction of gas removed from the
satellites as a function of the satellite galaxy's properties, distance to the
host and the strength of the feedback outflow. Applying these results to a
Milky Way like dark matter halo, we find that this singular shock ram pressure
stripping event may be quite efficient in removing the gas from the satellites
provided that they are closer than $\sim 50-100$ kpc to the host. We also use
the orbital and mass modelling data for eight Galactic dwarf spheroidal (dSph)
satellites, and find that it is likely that many of them have been affected by
the Galactic outflow, although the current data still leave much room for
uncertainties. Finally, we point out that galactic outflows of the host may
also trigger a starburst in the satellite galaxies by over-pressuring their
gas discs. We speculate that this process may be responsible for the formation
of the globular clusters observed in some of the Milky Way's dSphs (e.g. the
Fornax and Sagittarius dSphs) and may also be important for the formation of
the bulk stellar populations in the dSphs.
\end{abstract}


\section{Introduction}\label{intro}

A potentially important problem for the current cosmological galaxy formation
models was noted by \cite{KlypinEtal99} and \cite{MooreEtal99} in that the
number of observed dwarf galaxy satellites is too small by a factor of at
least ten compared with the simulations \citep[for a recent review,
  see][]{Bullock10}. The suggested solutions to the problem are
that dark matter halos of smaller mass are inefficient in acquiring their gas
\citep{BullockEtal00} or turning that gas into stars, being more easily
disrupted by star formation feedback \citep{DekelSilk86}. Alternatively, dwarf
galaxies may be susceptible to influences from their host galaxies, e.g., due
to tidal forces \citep{MayerEtal01} or by ram pressure stripping
\citep{MayerEtal06}.  In the last decade, a number of ultra-faint galaxies
were detected \citep[e.g.,][]{SimonGeha07}, firmly suggesting for the first
time that there is indeed a number of currently undetected almost baryon-free
dwarf galaxies,  supporting the idea that the astrophysical
  solutions listed above are responsible for modulating the numbers of
  satellite galaxies.

Interestingly, larger galaxies, including the Milky Way \citep{McGaughEtal09}
also appear to lack about half of their baryons compared with the universal
cosmological mass fraction of $\approx 0.16$ \citep{CenOstriker99}. The
missing gas does not seem to congregate in halos around the galaxies
\citep{AndersonBregman10}, probably having been ejected by AGN and star
formation feedback outflows well outside the galaxies
\citep{King05,DiMatteo05}.  For a recent detailed census of the baryons in
  the local Universe see \cite{ShullEtal11}.

The proposition that we make in this paper is that it is possible that the two
missing baryon problems are in fact causally related.  \cite{MayerEtal06} show
convincingly that ram pressure stripping, together with gravitational tides,
is able to remove a significant fraction of gas from dwarf spheroidal
galaxies (dSphs) orbiting the Milky Way. The ram pressure stripping is thus
found to be important despite the low present day density of gas in the
galactic halo \citep{AndersonBregman10}. Our key suggestion is that ram
pressure stripping during the short but intense galactic outflow phases could
be even more important because (a) the density of the outflowing gas, $\rho$,
could be much higher than that of the present day Galactic gaseous halo, and (b) the
outflow velocity, $V_{\rm sh}$, is also likely to be higher than the relative
velocity between the satellite and the hot {\em static} halo. Since the ram
pressure stripping is proportional to the ram pressure of the ambient gas,
$P_{\rm ram} = \rho V_{\rm sh}^2$, the efficiency of this transient gas
removal phase may be significant.

To demonstrate this point more clearly, we plot in Figure 1 the ram pressure
seen by a satellite galaxy for three different models. The ``Hot Halo'' model
(HH hereafter) follows \cite{MayerEtal06}, which assumes that the gas
temperature is equal to the virial temperature of the Milky Way's halo
\citep[modelled as an NFW potential, see][; halo parameters are listed in \S 3
  below]{NFW}, and the satellite's orbital velocity is equal to twice the
local circular speed. This is approximately correct at the pericentre of an
eccentric orbit, where \cite{MayerEtal06} find that most mass is stripped
off. 

The two other ``shock'' models both mimic the physics of the AGN feedback
theory by \cite{King03,King05}. This theory is based on observed fast ($v_{\rm
  AGN} \sim 0.1 c$) outflows from the nuclear (sub-parsec) regions of AGN
\citep{KP03}. In the inner fraction of a kpc, these outflows are in the
momentum-driven regime, which means that the fast outflow cools rapidly when
shocked, and thus only its momentum is used to drive the gas out of the host
galaxy. At larger radii that are of interest to us here, the fast outflow
shock becomes non-radiative, and its energy is retained in the shocked primary
outflow and in the kinetic energy of the shocked galaxy's shell
\citep{ZK12a}. The outflow is then powered mainly by the kinetic energy of the
fast wind, which is liberated at the rate $\dot E_{\rm AGN} \sim (v_{\rm
  AGN}/2c)\, L_{\rm Edd}$, where $L_{\rm Edd}$ is the Eddington luminosity for
the supermassive black hole launching the outflow.

In this model, in the spherically symmetric singular isothermal potential, the
shocked shell outflow velocity, $V_{\rm sh}$, is constant beyond $\sim$ a kpc
distance from the SMBH, and is a few times the velocity dispersion of the
singular isothermal sphere, $\sigma$ \citep[see figures in][]{KZP11}. This may
appear paradoxical at first as the mass of the shocked gas increases as the
AGN outflow sweeps through the galaxy. However, the energy and momentum of the
nuclear AGN outflow increase linearly with time as well, so the ratio of the
shocked mass to the fast AGN outflow mass actually stays constant, and this is
why the velocity of the shocked shell is constant.

This analytical model can be only approximately correct for a more complicated
non-spherically symmetric potential and gas distribution. It is clear that the
outflow velocity is smaller in directions where the gas density is larger,
e.g., along the galaxy midplane for a spiral galaxy \citep[cf. the arguments
  of][for the Fermi Bubbles in the Milky Way]{ZubovasEtal11a,ZN12a}. We do not
consider this level of detail in our exploratory models here; we simply choose
a constant outflow velocity in a spherically symmetric galaxy.  Thus, in our
model, the satellite ploughs through an outflow with velocity $V_{\rm sh} =
500$ km\,s$^{-1}$. We make two different assumptions for $M_{\rm sh}$, the
mass of the outflowing shell.  In the ``NFW shell'' model we assume that the
mass of the outflowing gas is equal to 0.05 times the enclosed total mass of
the halo at a given radius $R$. The mass of the outflow thus increases with
radius in this model.  The remaining ``$M_{\rm sh} =$~const'' model sets the
outflowing mass to $M_{\rm sh} = 5\times 10^{10}\msun$, independent of radius
$R$, which is 2.5 per cent of the host galaxy mass within its virial radius
(see section \ref{sec:real} for our Milky Way galaxy model).  The first
assumption is reasonable for a shell that is being continuously swept up and
so its mass increases as it travels outward \citep{KZP11}, whereas the second
assumption is reasonable if a significant fraction of the gas within the
virial radius first contracted to a high density central region e.g. the
proto-bulge) and was subsequently ejected by the AGN and stellar feedback.

We see that the ram pressure in the two shock models is much larger than that
in the present day hot halo model, which is mainly due to the fact that the
present day's Galactic gas halo is a low mass density one. The purpose of this
paper is to investigate in greater detail the ram pressure stripping of
smaller satellite galaxies by galactic outflows ultimately driven by SMBH and
starburst feedback in the parent galaxy. To this end we first consider a toy
model in which the galaxies are modelled as singular isothermal sphere (SIS)
potentials. The advantage of this model is that it is purely analytical and
easy to follow. We then consider a more realistic~\cite{NFW} dark matter
potential model for both the parent and the satellite, and assume that the
dwarf galaxy contains both a gaseous disk and a gaseous halo. We consider
separately the ram pressure effects on both these gas components.  Finally, we
apply the results to a sample of Milky Way dwarf galaxies.

\begin{figure}
\psfig{file=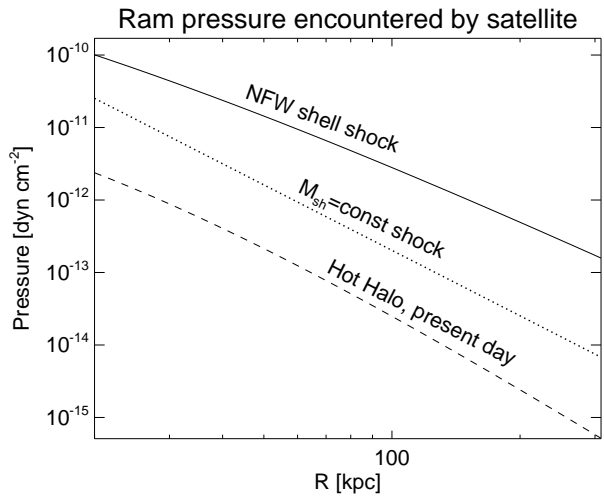,width=0.49\textwidth,angle=0}
\caption{The ram pressure sampled by a satellite galaxy in the present day
  Galactic hot halo (lower curve), compared to the ram pressure in the
  Galactic outflow at velocity $V_{\rm sh}=500$~km~s$^{-1}$, for two different
  assumptions about the mass in the outflow (see text for detail).}
\label{fig:continuous}
\end{figure}

\section{A toy spherical singular isothermal sphere model}\label{sec:toy}

\subsection{Model galaxies}

We start out by assuming that both the host and the dwarf satellite galaxies
can be modelled by fixed singular isothermal sphere potentials with circular
velocities $V_{\rm circ} = 200 v_{200}$ km~s$^{-1}$ and $v_{\rm circ} = 20
v_{20}$ km~s$^{-1}$, respectively. We assume that these potentials are
dominated by dark matter, and that the gas makes up a small fraction of the
total mass, so that the potential is approximately independent of the presence
of gas.

We further assume \citep{MMW98} that the virial radius of the halo of the model galaxies is given by 
\begin{equation}
R_{\rm vir} = {V_{\rm circ} \over 10 H(z)}\;,
\label{r200}
\end{equation}
where $H(z)$ is the Hubble constant at redshift $z$. The total mass within
the halo is related to $V_{\rm circ}$ as
\begin{equation}
M_{\rm tot} = {V_{\rm circ}^2 R_{\rm vir}\over G} = {V_{\rm circ}^3\over 10 G
  H(z)}\;.
\label{mvir}
\end{equation}

Consider a satellite orbiting the host galaxy a distance $R$ from the centre
of the host. The gas density profile inside the satellite is assumed to follow
the dark matter with a constant fraction $f_{\rm d} < 1$ of the dark matter
density. The gas surface density at radius $r$ from the centre of the
satellite, projected along the direction from the centre of the host, is
\begin{equation}
\Sigma_{\rm g}(r) = {f_{\rm d} v_{\rm circ}^2 \over \pi G r}\;.
\label{sigmad}
\end{equation}

The gas in this model dwarf galaxy is acted upon by a massive rapidly moving
shell of gas ejected from the parent galaxy, with the mass $M_{\rm sh}$ and
velocity $V_{\rm sh}\gg v_{\rm circ}$. The column density of the shell is
\begin{equation}
\Sigma_{\rm sh} = {M_{\rm sh}\over 4\pi R^2} = {f_g V_{\rm circ}^2 \over 4\pi G R}\;,
\label{sigma_sh}
\end{equation}
where $f_g\ll 1$ describes the mass of the ejected shell with respect to the
enclosed total mass at radius $R$ of the parent galaxy. The density of the
shell is
\begin{equation}
\rho_{\rm sh} = {M_{\rm sh}\over 4\pi R^2 \Delta R}\;,
\label{rho_sh}
\end{equation}
where $\Delta R < R$ is the radial thickness of the incoming shell. The shell
thickness is given by the difference between the forward shock velocity and
that of the contact discontinuity, $V_{\rm cd}$ \citep[see Fig. 1
  in][]{ZK12a}. For the specific heat ratio of $\gamma = 5/3$ for the ambient
shocked gas, those authors found that the result is $\Delta R = R/3$.

\subsection{Shock ram pressure stripping}\label{sec:ram}

Following the standard \citep{GunnGott72}, but at best approximate
\citep[see][]{MayerEtal06}, treatment of ram pressure stripping, we assume
that gas is ejected at radii where the restoring force per unit area, $\sim
2\pi G \Sigma_{\rm g} \Sigma_{\rm dg}$, is smaller than the ram pressure of
the shell's material, $P_{\rm sh} = \rho_{\rm sh} V_{\rm sh}^2$
\citep{GunnGott72}. Here $\Sigma_{\rm dg}$ is the total surface density of the
dwarf galaxy (including the Dark Matter; the gas only surface density is named
$\Sigma_{\rm g}$). In this regime, the shock stripping radius, defined by
\begin{equation}
P_{\rm sh} = 2\pi G \Sigma_{\rm g}(r_S) \Sigma_{\rm dg}(r_S)\;,
\label{rsh2a}
\end{equation}
 is equal to
\begin{equation}
r_S = \left({8 f_{\rm d} v_{\rm circ}^4 R^3 \over 3 V_{\rm sh}^2 G M_{\rm
    sh}}\right)^{1/2}\;,
\label{rs3}
\end{equation}
where we assumed that $\Delta R = R/3$. Using the approximation $GM_{\rm
  sh}/ R = f_g V_{\rm circ}^2$, we have
\begin{equation}
r_S = \left({8 f_{\rm d}\over 3 f_g}\right)^{1/2} {v_{\rm circ}^2 R\over
  V_{\rm circ}
  V_{\rm sh}} \approx 0.65 \hbox{ kpc}  {v_{20}^2 R_{100}\over V_{200}
  V_{500}} \left({f_{\rm d}\over
  f_g}\right)^{1/2}\;,
\label{rs4}
\end{equation}
where $R_{100}=R/100$~kpc and $V_{500} = V_{\rm sh}/(500$ km~s$^{-1}$).  The
value of $r_S$ given by this equation is substantially smaller than the virial
radius, $r_{\rm vir}$, for the dwarf galaxy, indicating that most of the gas
would be shock-stripped in this simple model, as we now show.

We can now calculate the fraction of the dwarf's gas mass that is retained
after the passage of the host's feedback outflow, $\delta_{\rm
  ret}$. Evidently, since the enclosed gas mass at radius $r$ is directly
proportional to $r$ in our toy SIS potential model, $\delta_{\rm ret} =
r_S/r_{\rm vir}$, where $r_{\rm vir}$ is given by equation \ref{r200} with the
circular velocity appropriate for the dwarf galaxy, and $r_S$ is calculated
above. Based on equation \ref{rs4}, the result is:
\begin{equation}
\delta_{\rm ret} = \sqrt{8 f_{\rm d}\over 3 f_g} \;{v_{\rm circ} \over V_{\rm
    sh}}\; {R \over R_{\rm vir}} \;.
\label{delta1}
\end{equation}
For $f_{\rm d} \sim f_g$, this predicts almost a complete loss of gas from the
dwarf galaxy anywhere inside the host halo, e.g., $R < R_{\rm vir}$.

It is instructive to compare this shock stripping mechanism with that due to
gravitational tides within the halo of the galaxy host. Within the singular
isothermal potential approximation for both the host and the satellite and the
mass-radius scaling relations given by equations \ref{r200} and \ref{mvir},
the density of a galaxy at its effective radius is independent of the galaxy's
mass or circular velocity. Therefore the dwarf galaxy would also be tidally
stripped in this model if it fell inside the host's halo. Equation
\ref{delta1} is thus somewhat academic.

However, outside the host halo, the SIS model predicts no tidal destruction
for the dwarf, whereas the shock stripping mechanism is still effective out to
a radius several times $R_{\rm 200}$. This toy model is clearly very
over-simplified compared with realistic galaxies, but it does indicate that
the effect may be large, and calls for a more detailed investigation which we present below.

\section{A more realistic model}\label{sec:real}

We now build a slightly more realistic model for both the host and the dwarf
galaxy by using the \cite{NFW} potentials for the dark matter halos and the
\cite{MMW98} model for the disc of the dwarf galaxy. For definiteness, we
set the host's virial mass to $M_{\rm host} = 2\times 10^{12}\msun$, which
gives a virial radius of $R_{\rm vir} = 204$~kpc, and a circular velocity
at the virial radius of 205 km~s$^{-1}$. We set the concentration parameter to
$c=20$. This host galaxy model is compatible with the DM halo of the Milky Way~\citep[see e.g.][]{Gnedin2010}.

For the dwarf galaxy, we consider two plausible gas distributions within its
dark matter halo: one distributed as the dark matter, with gas mass fraction
$f_h = 0.1$, and the other sitting in a rotation supported disc with the mass
fraction of $f_{\rm d} = 0.05$. The disc surface density follows an
exponential density profile \citep[see][]{MMW98} with the scale-radius given
by $r_{\rm d} = 0.05 r_{\rm vir}$, where $r_{\rm vir}$ is the virial radius
for the dwarf galaxy. We assume that the concentration parameter of the
\cite{NFW} halo, $c_{\rm NFW} = 10$ for the dwarfs. This determines the
potentials of the dark matter profiles and the disc in the dwarf galaxy
completely; we neglect the contribution to the potential due to the disc of
the host galaxy.

We should also specify the mass of the shell being expelled from the host,
$M_{\rm sh}$, which may be a function of the galactocentric radius, and the
velocity of the shell, $v_{\rm sh}$. To sample the range of possible outcomes
we test two opposite assumptions for $M_{\rm sh}$ as explained in the
Introduction: (i) that the shell mass $M_{\rm sh} = f_{\rm sh} M_{\rm
  enc}(R)$, where $f_{\rm sh}=0.05$ and $M_{\rm enc}(R)$ is the total mass
enclosed within radius $R$, and (ii) that $M_{\rm sh} = 5\times 10^{10}\msun$,
independent of radius $R$.

Figure \ref{fig:continuous} shows the fraction of gas retained in the dwarf
galaxy after the host feedback shock passage for the disc and the halo
components, and also for the total gas mass, naturally defined as the sum of
the halo and the disc masses. Three different total masses for the dwarf are
considered, resulting in three different values for the maximum circular
velocity in the galaxy, as marked in the top left corner of each panel.  All
the black curves in the figure are for the default shock velocity studied in
this paper, $V_{\rm sh} = 500$ km~s$^{-1}$. We see that the halo gas component
of the dwarf is the one easiest to remove as it is comparatively more extended
than the exponential disc. The disc component is the most compact one, thus a
$\sim 90$\% removal of the disc requires the smallest dwarf (the top panel) to
be at $R\simlt 100$ kpc from the centre of the host. The figures show that the
more massive the dwarf galaxy (i.e. the larger its $v_{\rm circ}$) the harder
it is to affect its gas by the shock from the host, since all the curves shift
to smaller radii as we compare the top panel to the bottom one. Note that
these trends are qualitatively consistent with the toy SIS model, cf. equation
\ref{delta1}. The latter predicts that the radius at which a given fraction of
gas is removed scales inversely with $v_{\rm circ}$, which is approximately
borne out in Figure \ref{fig:continuous}. For example, the $\delta_{\rm ret} =
0.1$ for the disc is reached at $R\approx 30$ kpc for the $v_{\rm circ} = 60$
km~s$^{-1}$ versus $R\approx 100$ kpc for the $v_{\rm circ} = 15$ km~s$^{-1}$.

To estimate the sensitivity of our models to the assumed value of $V_{\rm
  sh}$, we also computed the fraction of gas retained in the halo of the
satellite galaxy for two other values of $V_{\rm sh}$, e.g., 300 and 1000
km~s$^{-1}$ for the blue and the red curves, respectively, for the middle
panel. We see that the gas is removed from the satellite more efficiently by a
faster shock, as should be expected. Qualitatively, the dotted curve appears
to shift to larger radii in roughly linear proportionality to $V_{\rm
  sh}$. For example, the radius at which 90\% of the halo is removed moves
from $\sim 100$~kpc to $\sim 300$~kpc as $V_{\rm sh}$ is changed from 300 to
1000 km~s$^{-1}$.

Figure \ref{fig:m5} shows the same calculation but now for the fixed mass of
the expelled shell (case (ii) above). We observe that because of the $\sim
1/R^3$ ram pressure fall in this model, the transition from the strongly
affected, e.g., $\delta_{\rm ret}\ll 1$, to the weakly affected regime,
$\delta_{\rm ret}\sim 1$, occurs over a more restricted range of radii than in
Figure \ref{fig:continuous}. Furthermore, since the ram pressure in this model
is lower (see Figure 1), the satellite galaxies need to be even closer to the
centre of the host to be affected by the feedback shock.

\begin{figure}
\psfig{file=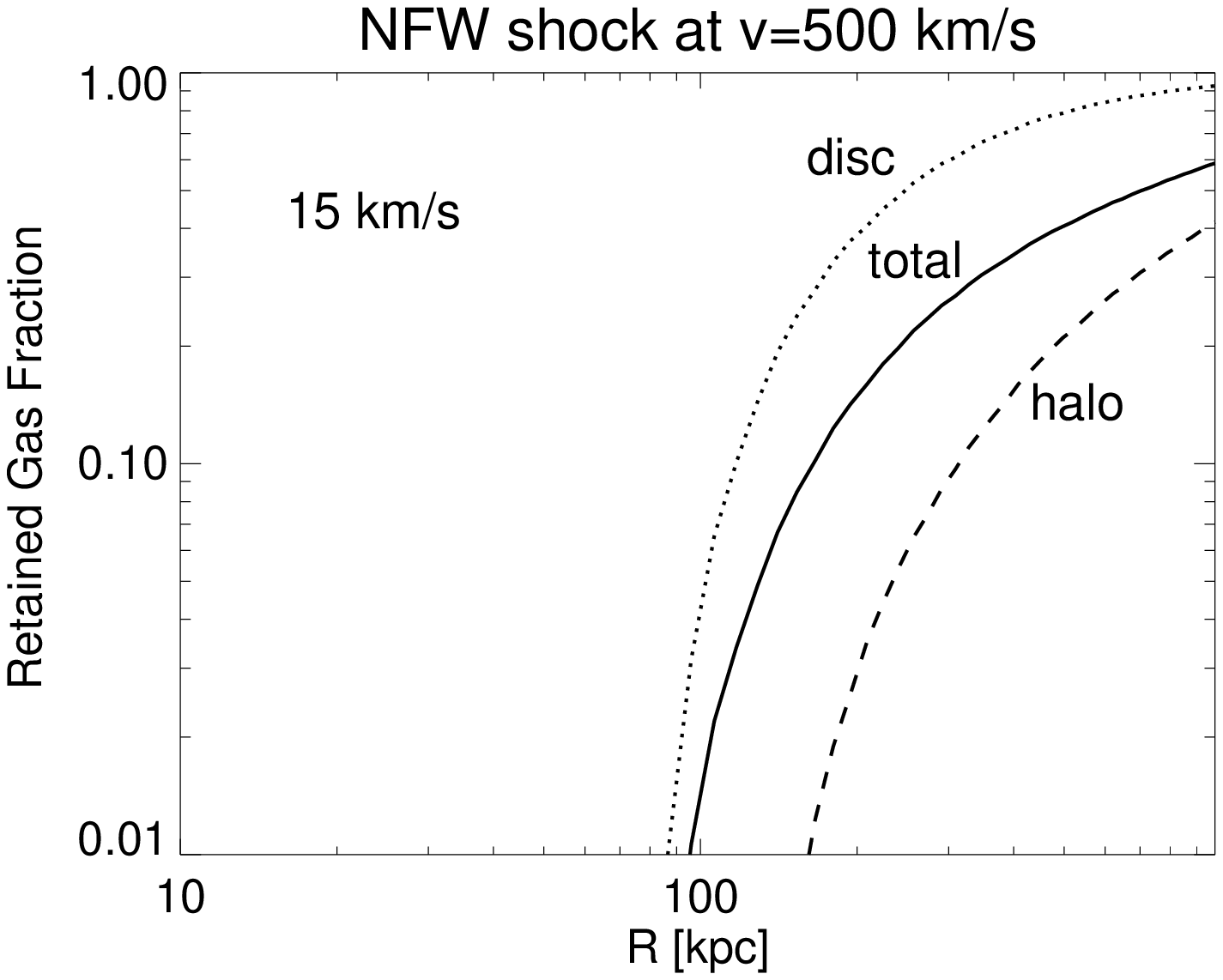,width=0.45\textwidth,angle=0}
\psfig{file=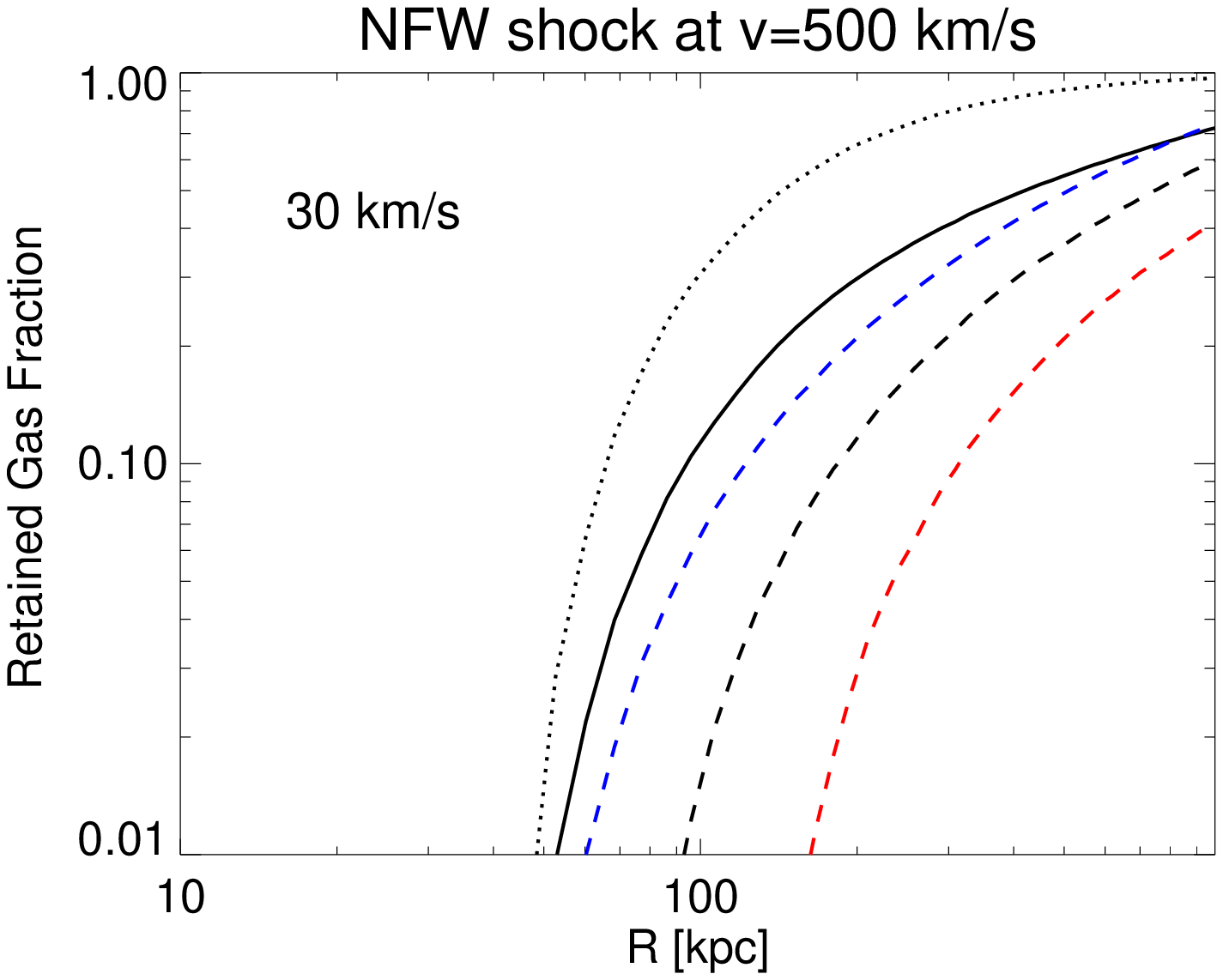,width=0.45\textwidth,angle=0}
\psfig{file=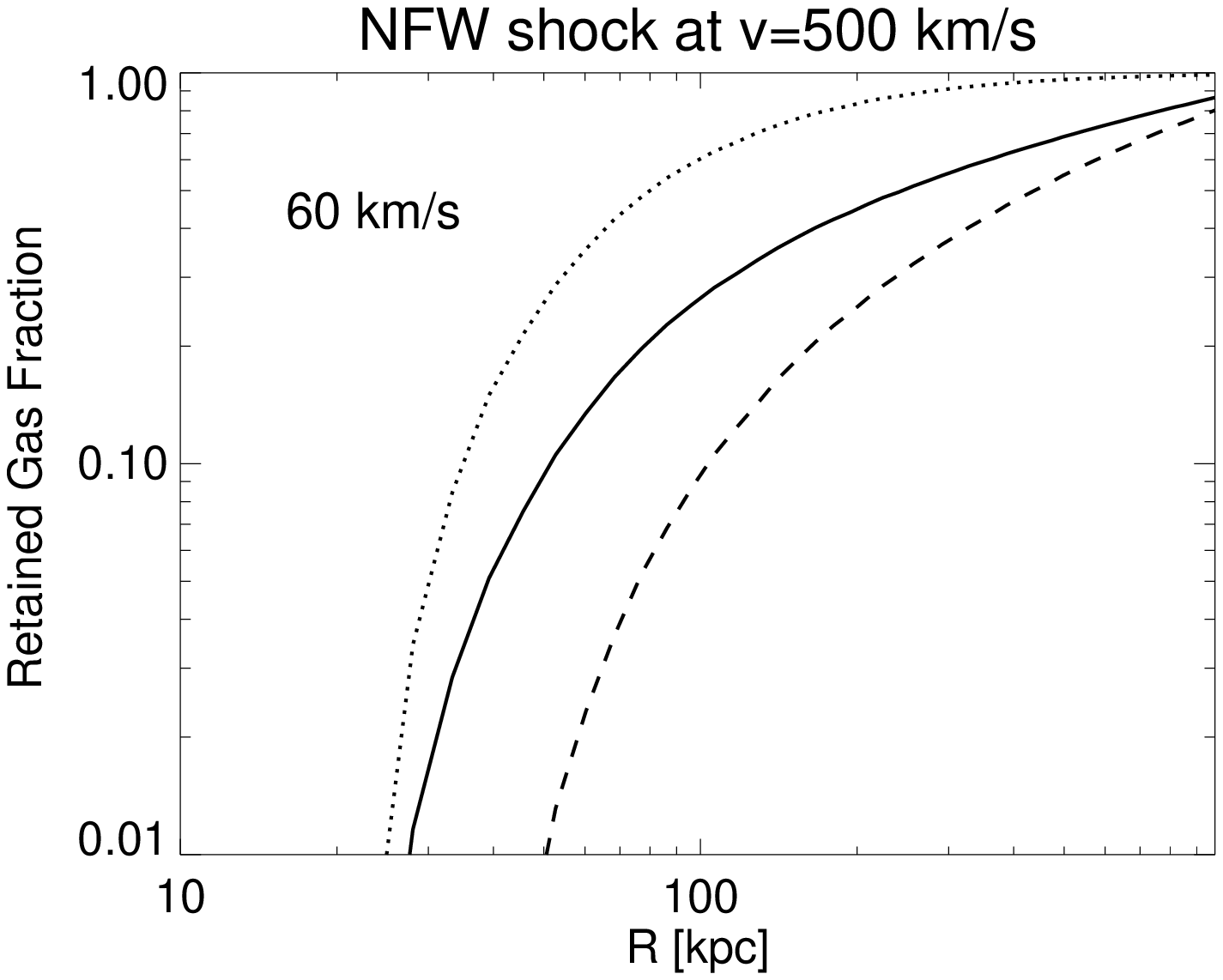,width=0.45\textwidth,angle=0}
\caption{Fractions of gas retained in the dwarf galaxy's disc and halo after
  passage of the host galaxy's shock, as well as the total fraction of gas
  retained. Components are indicated next to the respective curves, as
  functions of the galactocentric distance, $R$. The dwarf galaxies considered
  have three different virial masses, corresponding to the maximum circular
  velocities of 15, 30 and 60 km~s$^{-1}$, from top to bottom,
  respectively. The calculation assumes that the shell mass increases as it
  propagates outward (see text in \S \ref{sec:real}). All the black
    curves are for the shock velocity of $V_{\rm sh} = 500$~km~s$^{-1}$,
    whereas the blue and the red are for $V_{\rm sh} = 300$~km~s$^{-1}$ and
    $V_{\rm sh} = 1000$~km~s$^{-1}$, respectively.}
\label{fig:continuous}
\end{figure}

\begin{figure}
\psfig{file=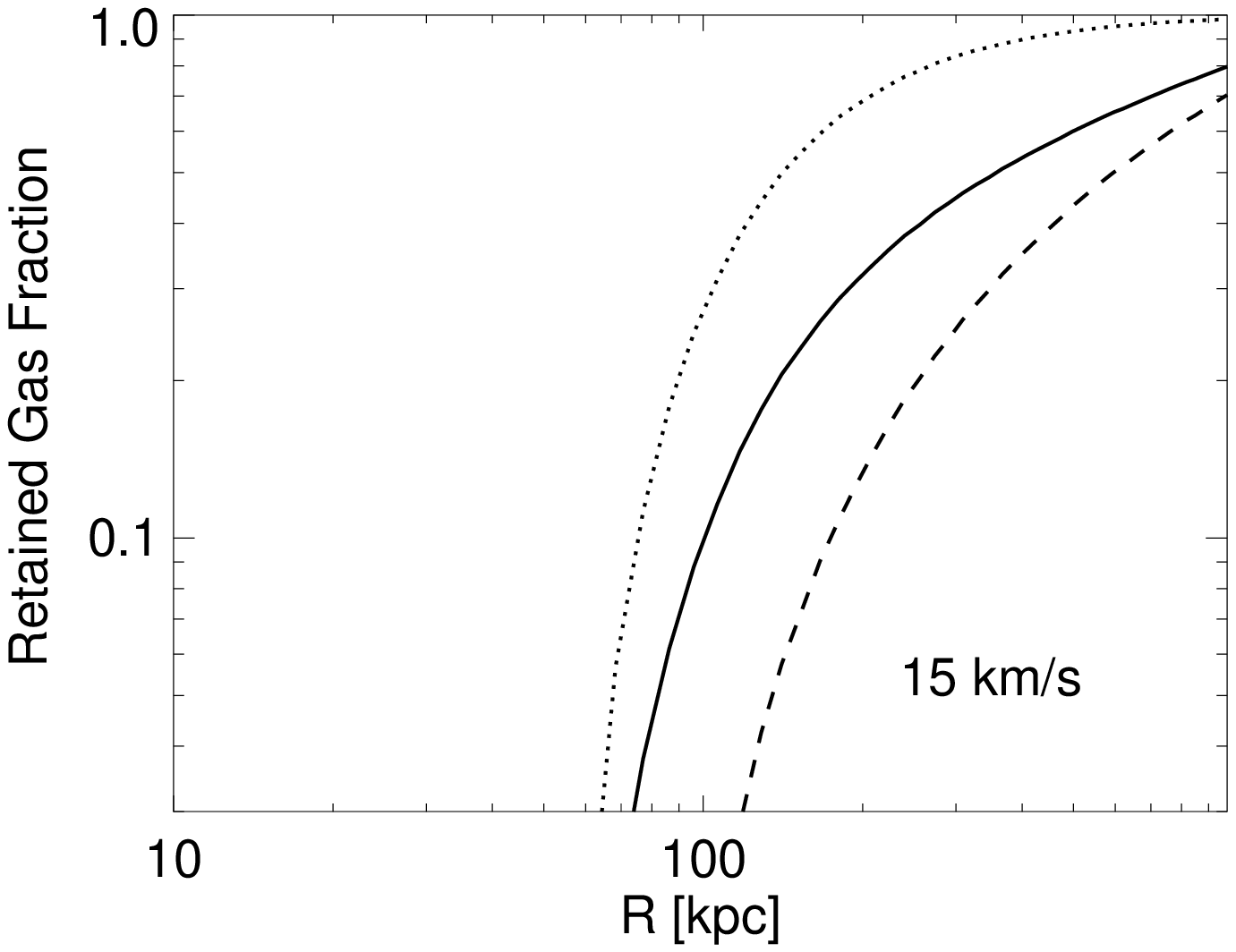,width=0.45\textwidth,angle=0}
\psfig{file=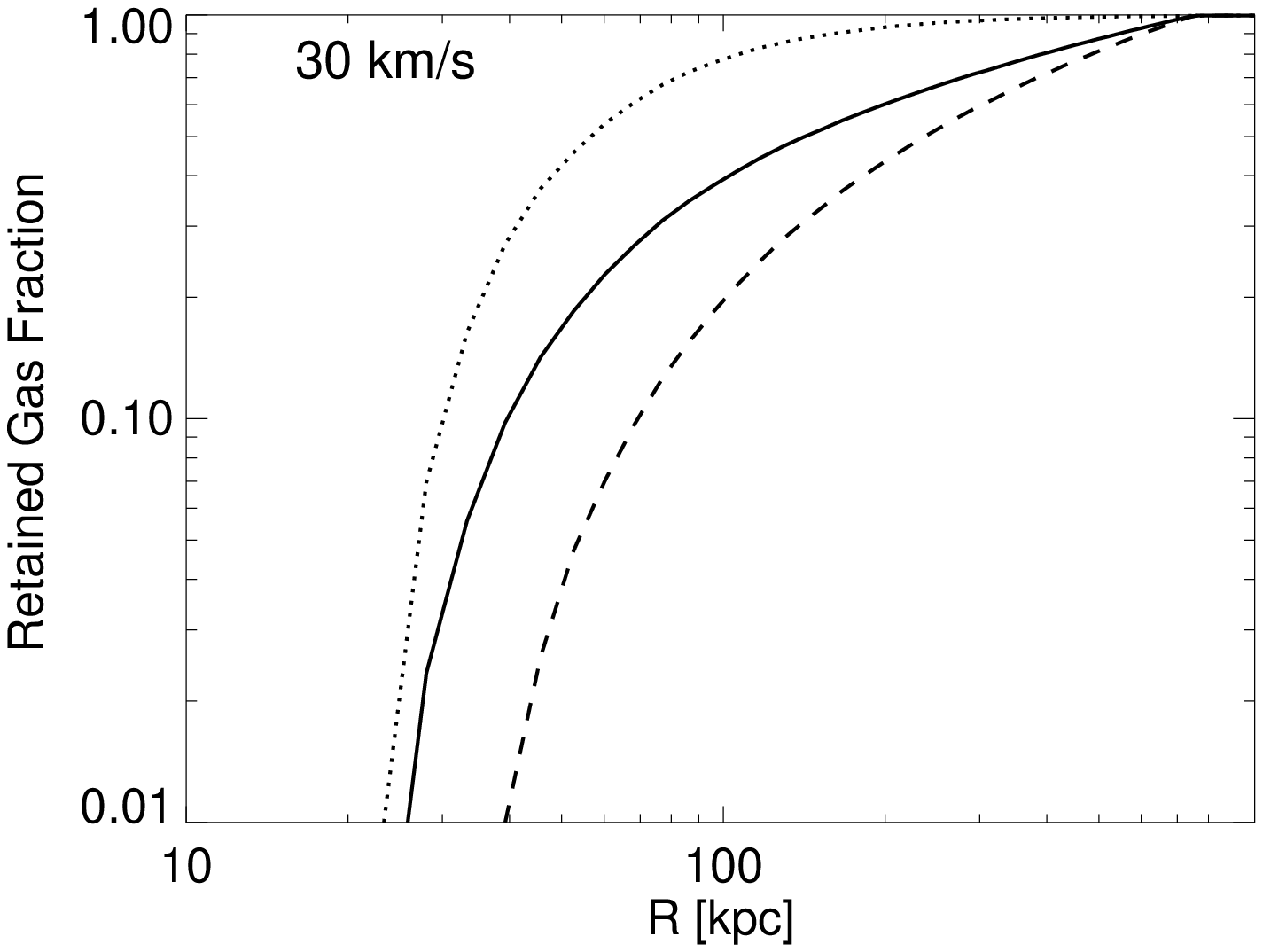,width=0.45\textwidth,angle=0}
\psfig{file=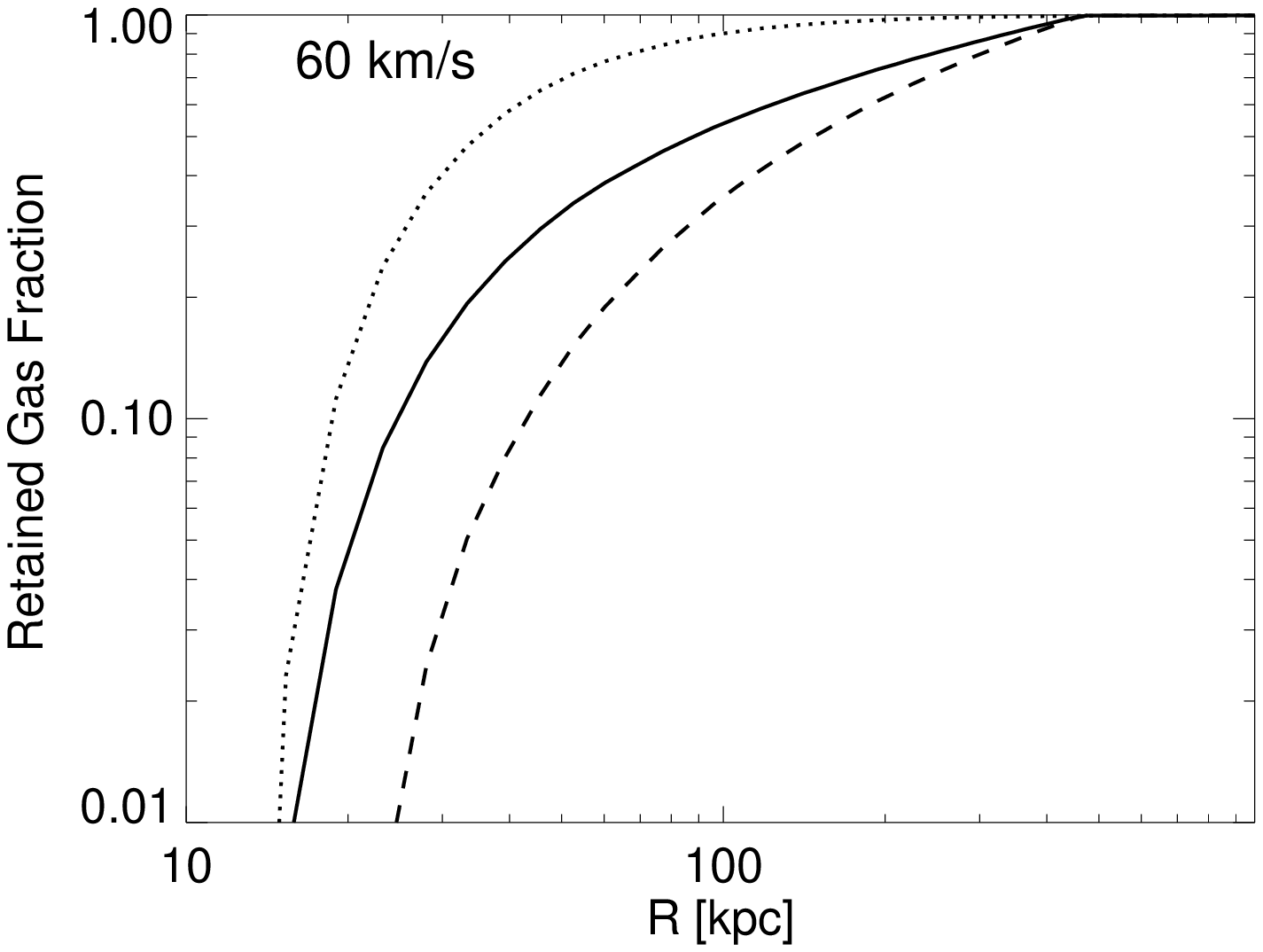,width=0.45\textwidth,angle=0}
\caption{Same as Figure \ref{fig:continuous} but for a fixed mass in the
  expelled shell of $M_{\rm sh} = 5\times 10^{10}\msun$. The meaning of the
  solid, dotted and dashed curves is the same as in Figure
  \ref{fig:continuous}. }
\label{fig:m5}
\end{figure}

\section{Application to the Milky Way dSphs}\label{sec:mw}

We now compare the results of our simple calculations with
  the observed data for the dwarf spheroidal (dSph) satellites of the
  Milky Way. For more than a decade, a concerted observational effort
  has been underway to determine the dark matter content of the Milky
  Way dSph population~\citep[see ][for a recent review]{Walker2012}
  due to their importance for understanding galaxy formation on small
  scales. Although data are now available for more than twenty dSphs,
  in what follows we consider only the so-called ``classical'' dSphs,
  as these more luminous objects have constraints on both their dark
  matter content~\citep{Walker2009} and Galactocentric
  orbits~\citep{Lux2010}.

  Table~\ref{tab:mcmcresults} presents the data we have used in our
  comparison. The mass estimates at $r_{\rm half}$ and $r_{\rm last}$
  are taken from~\cite{Walker2009} who used Jeans equation modelling,
  combined with a Markov-Chain-Monte-Carlo (MCMC) algorithm, to determine
  masses for the dSphs based on their projected velocity dispersion
  profiles. We take the values for the orbital apo- and peri-centres
  to be those obtained by~\cite{Lux2010} who applied an MCMC approach
  to the modelling of the space motions of the six dSphs which have
  been the subject of long-term proper motion studies.

\begin{table*}
\begin{center}
\scriptsize
\begin{tabular}{lclrrccrrr}
\\\hline\hline Object &L$_V$ / L$_{V,\odot}$&$r_{\mathrm{half}}$ / pc&$d$ /
kpc&$r_{\rm last}$ / kpc&M$(r_{\rm half})$ / $10^7$M$_\odot$&M$(r_{\rm last})$
/ $10^7$M$_\odot$&$f_{\rm b,last}$&$r_{\rm
  peri}$ / kpc & $r_{\rm apo}$ / kpc\\ \hline

Carina    &(2.4 $ \pm $1.0)$\ \times$10$^5$&241$\pm$23&101$\pm$5&0.87&$0.4_{-0.1}^{+0.1}$&$3.7_{-1.8}^{+2.1}$&0.006&$ 60 \pm 30 $&$ 110 \pm 30 $ \\ \hline
Draco     &(2.7 $ \pm $ 0.4)$\ \times$10$^5$&196$\pm$12&82$\pm$6&0.92&$0.6_{-0.3}^{+0.5}$&$26.4_{-17.4}^{+18.6}$&0.001&$ 90 \pm 10 $&$ 300 \pm 100 $ \\ \hline
Fornax   &(1.4 $ \pm $ 0.4)$\ \times$10$^7$&668$\pm$34&138$\pm$8&1.7&$4.3_{-0.7}^{+0.6}$&$12.8_{-5.6}^{+2.2}$&0.13&$ 120 \pm 20 $&$ 180 \pm 50 $ \\ \hline
Leo I      &(3.4 $ \pm $ 1.1)$\ \times$10$^6$&246$\pm$19&250$\pm$30&0.93&$1.0_{-0.4}^{+0.6}$&$8.9_{-5.2}^{+4.3}$&0.04& $-$ & $-$  \\ \hline
Leo II     &(5.9 $ \pm $ 1.8)$\ \times$10$^5$&151$\pm$17&205$\pm$12&0.42&$0.5_{-0.3}^{+0.2}$&$1.7_{-1.2}^{+1.9}$&0.03&$ -  $&$ - $ \\ \hline
Sculptor &(1.4 $ \pm $ 0.6)$\ \times$10$^6$&260$\pm$39&79$\pm$4&1.1&$1.0_{-0.3}^{+0.3}$&$10.0_{-5.0}^{+3.2}$&0.01&$ 60 \pm 10 $&$ 160 \pm 80 $ \\ \hline
Sextans  &(4.1$ \pm $ 1.9)$\ \times$10$^5$&682$\pm$117&86$\pm$4&1.0&$1.6_{-0.4}^{+0.4}$&$2.0_{-0.7}^{+1.0}$&0.02&$ 70 \pm 20 $&$ 300 \pm 200 $ \\ \hline
Ursa Minor&(2.0 $ \pm $ 0.9)$\ \times$10$^5$&280$\pm$15&66$\pm$3&0.74&$1.3_{-0.5}^{+0.3}$&$4.4_{-2.0}^{+2.9}$&0.005&$ 40 \pm 20 $&$ 90 \pm 20 $ \\ \hline

\end{tabular}
\caption{Table of dSph structural parameters and results of mass
  modelling. The columns are: (1) dSph name; (2) observed V-band
  luminosity~\protect\citep[from ][]{IH95}; (3) projected half-light
  radius (the radius enclosing half of the total luminosity), as listed by~\protect\citeauthor{Walker2009}~\citeyear{Walker2009},
    assuming spherical symmetry and
    a~\protect\citeauthor{King1962}~\citeyear{King1962} surface brightness
    profile with parameters taken
    from~\protect\citeauthor{Mateo1998}~\citeyear{Mateo1998}; (4) Distance~\protect\citep[from ][]{Mateo1998};(5) the
  radius of the outer bin in the velocity dispersion profile used in the mass
  modelling of~\protect\cite{Walker2009}; (6) total mass inside $r_{\rm
    half}$~\protect\citep{Walker2009}; (7) total mass inside $r_{\rm
    last}$~\protect\citep{Walker2009}; (8) baryon fraction inside $r_{\rm
    last}$; (9 ,10) orbital pericentre and apocentre estimated
  by~\protect\cite{Lux2010} assuming the Milky Way halo model
  of~\protect\cite{Law2005}. Note: In calculating the stellar masses from the
  measured luminosities, we assume that M/L$_{\rm V}$=1 for the stellar
  components of the dSphs, and that the entire luminosity is contained within
  the radius $r_{\rm last}$. The latter assumption is likely not valid for
  Sextans.}
\label{tab:mcmcresults}
\end{center}
\end{table*}

\begin{table*}
\begin{center}
\scriptsize
\begin{tabular}{lccccccr}
  \\\hline\hline Object &$r_{\rm d}$, kpc &$M_{\rm 0}$ & HH  & A & A$_{\rm max}$ & A$_{\rm min}$ & $M_{\rm obs}$ \\ \hline

Carina   &0.56 &16.6  &5.2     &0.74      &2.39   &0.0016   &0.24 \\ \hline
Draco    &1.47$^1$ &302.4 &248.3   &251.0    &268.4  &212.9    &0.27 \\ \hline 
Fornax   &0.75 &40.3 &31.0     & 16.4    & 21.4  &9.0    &14.0  \\ \hline
Leo I    &0.84  &55.4 &52.4     &34.1    & 42.6  &13.7   &3.4  \\  \hline
Leo II   &0.55  &15.6 &  13.5    &5.3    &8.7    &0.32   &0.59  \\ \hline
Sculptor &0.82  &52.9 & 25.0     &20.1   &31.0   &2.37   &1.4 \\ \hline
Sextans  &0.39  &5.5 &   1.32    &2.3    &3.69   &$2.5\times 10^{-5}$ &0.41  \\ \hline
Ursa Minor &0.66 &27.7 &4.5      &0.79   &2.64   &0.005           &0.2   \\ \hline

\end{tabular}
\caption{Results of our ram pressure stripping modelling for the dSphs from
  Table 1. The columns are: (1) dSph name; (2) the disc scale radius for the
  model dwarf galaxy; (3) the initial mass of the gas disc, in units of
  $10^6\msun$; (4) the mass of the gas retained in the HH model; (5) same for
  the NFW shock model with the dSph located at the estimated apocentre of the
  orbit, $a= a_0$ as in table 1; (6) same as (5) but for $a=a_0+ \delta a$;
  (7) same as (5) but for $a=a_0- \delta a$; (8) the observed stellar mass in
  the dSph.  Notes: $^1$The large estimate for $r_{\rm d}$ in the case of
  Draco arises due to the large value of M$(r_{\rm last})$ obtained by
  \protect\cite{Walker2009}. Those authors note that M$(r_{\rm last})$
  estimates are more model dependent than those of M$(r_{\rm half})$ - it is
  therefore possible that the mass of the NFW model for Draco that we use here
  is over-estimated. However, it serves to illustrate the impact of an outflow
  on a dSph with a more massive halo.}
\label{tab:shock_results}
\end{center}
\end{table*}

Our comparison has two main steps to it: (i) first we build a NFW dark matter
halo model and a corresponding gas disc model for each of the dSphs in
Table \ref{tab:mcmcresults}, and (ii) we determine the ram pressure acting on
the gas disc in the dSph and calculate how much of the gas remains in the
disc after the shock's passage. The results are summarised in Table 2.

To accomplish the first step, we calculate the circular velocity at $r=r_{\rm
  last}$ for the dSph, and assume that it corresponds to the maximum circular
velocity of a NFW halo hosting it.  Since the dSphs' circular velocity
profiles are rather flat in the interesting range of radii in the dSph
\citep[e.g., see Fig. 1 in][]{MayerEtal06}, this procedure does not appear to
introduce large uncertainties. We then insert a gas disc with properties as
described in \S \ref{sec:real}, following the model of \cite{MMW98}. The gas
mass of the model disc is given in column 3 of Table
\ref{tab:shock_results}. This mass is the initial mass of the disc before the
satellite is exposed to the influence of the ram pressure stripping.

For the second step, there is a range of possible models and further parameter
choices to make. As our analytical study is clearly a rough approximation
only, we calibrate its potential importance somewhat by applying the same
analytical formalism to the ram pressure stripping by the present day ``hot
halo'' (HH) model, which was shown to be effective for satellites on orbits
with pericentres of $\sim 50$ kpc by the numerical simulations of
\cite{MayerEtal06}.

\cite{MayerEtal06} show that the gas stripping effect is strongly maximised
near the pericentre of the satellite's orbit, which is natural as both the hot
halo density and the relative velocity of the galaxy and the gas reach their
peak values at that point. Therefore, for the HH model, we repeat the
procedure outlined in \S \ref{sec:ram} for determining the radius $r_S$
outside of which the gas disc of the dSph is stripped, but using the
model ram pressure appropriate for the Hot Halo. The mass of the gas retained
in the dSph in this model is shown in the fourth column of Table
\ref{tab:shock_results}.

Now, considering the effect of the galactic outflow on the satellite galaxies,
it is most realistic to assume that these were near apocentres of their orbits
at the time of the shock passage. This is because the objects spend most of
the time there, and the shock passage is most likely to have found them at
those more remote parts of their trajectories.

Table 1 shows that while the data on the dSphs have improved significantly
over the last 5 years or so, the orbits of the satellites are still rather
uncertain. In particular, the apocentres for many dSphs are uncertain by a
large factor, and are not known at all for Leo I and Leo II. To take these
uncertainties into account to some degree, we merely use the nominal, the
minimum and the maximum values of the apocentres from Table 1. That is,
denoting the apocentre values from Table 1 as $a_0\pm \delta a$ for each dSph,
we calculate its disc disruption by the shock for the three values of $a=
a_0$, $a= a_0 + \delta a$ and $a= a_0 - \delta a$, respectively, and list them
in columns marked by ``A'', ``A$_{\rm max}$'' and ``A$_{\rm min}$''. For Leo I
and II we assumed, somewhat arbitrarily, that their apocentres are equal to
their current galactocentric distance, e.g., $a_0 = d$, and that the error in
their apocentres is half that ($\delta a = d/2$).

Comparison of the ``HH'' and the ``A'' models from Table
\ref{tab:shock_results} demonstrates that for most dSphs the shock ram
pressure stripping is roughly as important as the hot halo ram pressure
stripping, although some dSphs are dominated by one effect or the other.  This
is somewhat counter-intuitive given the much higher ram pressure of the shock
outflow (Figure 1), but can be reconciled with the expectations by noting that
the satellites may be much further from the host in the model A than they are
in the model HH.

Further, the next two columns in Table 2 show that the present orbital data on
dSphs are still not sufficiently accurate to make a firm decision on whether
the feedback outflow stripping of the {\em observed} dSphs is important or
not. Indeed, column 6 (labeled ``A$_{\rm max}$'') shows that by placing the
  dSphs at the maximum distance consistent with the data, the effect
  of the shock passage is minimised to the point that only Carina and Ursa
  Minor are strongly affected. On the other hand, if the satellites are placed
  at the minimum allowed distance (column 7 labeled ``A$_{\rm min}$''), the
    gas stripping is strongly enhanced; only Draco remains largely unaffected.

In conclusion, there is clearly scope for the observed dSphs to have
suffered major disruption due to outflows from the Milky Way. However,
better orbital data and internal models for the dSphs are needed to
investigate this further.

\section{Induced star formation in the dSphs}\label{sec:induced}

There is another potentially significant way in which galactic
outflows could influence the observed properties of the Milky Way dSphs.

Massive stars are observed \citep{DeharvengEtal05} to produce not only
negative but also positive feedback on their immediate gaseous
environment. Shocks driven by supernova explosions, stellar winds
  or photo-ionisation can pressurise the surrounding ambient gas to high
  densities and result in star formation. Galactic outflows are very likely
to result in triggered star formation as long as the shocked ambient gas is
able to cool rapidly (Nayakshin \& Zubovas 2012, submitted).

We shall now show that the host galaxy outflow is capable of inducing star
formation in the dSphs. The mid-plane density in the gas disc of the dSph is
of the order of $\Sigma_{\rm g}/h$, where $h$ is the vertical scale height of
the gas disc, and $\Sigma_{\rm g}$ is the gas only surface density of the
dSph. Therefore, the disc pressure is
\begin{equation}
P_{\rm disc} \approx {\Sigma_{\rm g}\over h} c_s^2\;,
\end{equation}
where $c_s$ is the sound speed in the disc. For a gas disc in hydrostatic
balance, $c_s/v_{\rm circ} = h/r$ \citep{Shakura73}. Further, we approximate
$M_{\rm dg}(r) \approx \Sigma_{\rm dg}(r) \pi r^2$, where $M_{\rm dg}(r)$ is
the total enclosed mass within $r$ for the dSph, and $\Sigma_{\rm
  dg}(r)$ is the total surface density at that radius. Noting that $v_{\rm
  circ}^2 = GM_{\rm dg}/r = \pi G r \Sigma_{\rm dg}(r)$, we estimate the mid-plane
pressure in the gas disc as
\begin{equation}
P_{\rm disc} \sim {\Sigma_{\rm g} c_s^2\over h v_{\rm circ}^2} v_{\rm
  circ}^2\sim 2 \pi G \Sigma_{\rm d} \Sigma_{\rm dg} {h\over 2 r}\;.
\label{pdisk1}
\end{equation}
Regions where the disc pressure is smaller than the ram pressure of the
outflow, i.e., where
\begin{equation}
P_{\rm sh} > P_{\rm disc} = 2\pi G \Sigma_{\rm d}(r) \Sigma_{\rm dg}(r) {h\over 2 r} \;,
\label{rstar1}
\end{equation}
are susceptible to compression and triggered star formation. Comparing this
condition with a similar condition for removal of the gas from the dSph (disc
truncation) given by equation \ref{rsh2a}, we see that the zone where the
host's outflow can induce star formation in the satellite is larger. Indeed,
since $h/r<1$, there is a region of the disc inward from the shock truncation
radius, $r_S$, where $P_{\rm disc} < P_{\rm sh} < 2\pi G \Sigma_{\rm d}
\Sigma_{\rm gd}$, so that the gas in the disc of the dSph is compressed but
not expelled. At radii $r>r_S$ the gas is both expelled and compressed to
higher densities.

This implies that the host's galactic outflow can also trigger a local
starburst in the dSph satellite. Furthermore, there may be several mass
ejection episodes for the host galaxy, e.g., one during the birth of the
protogalaxy, and then more during major merger(s) powerful enough to trigger
either a quasar-driven outflow or a significant starburst in the host. As the
dSph presumably contains less and less gas as time progresses, it becomes
progressively easier to induce a starburst in it in each subsequent episode of
the host galaxy outflow (Eq.~\ref{pdisk1}). It is also worth noting that the
inner edge of the region in which star formation can be triggered moves
inwards as the gas mass in the dwarf decreases. Combined with the pollution of
the remaining gas by previous bursts of star formation this could provide a
natural explanation of the radial metalicity gradients observed in many
dSphs, in which more metal rich, and younger, populations are more centrally
concentrated~\citep[see e.g.][]{deBoer2012a,deBoer2012b}.  On the other hand,
shock gas stripping alone produces a metalicity gradient, since stars in the
outer regions of the dwarf galaxy may form only early on, before the gas was
expelled from those regions; the inner regions of the satellite can continue
star formation since they retain their gas.

Nayakshin \& McLaughlin (2013, submitted) argue that galaxy outflows produce
enormous pressures. In the case of quasar-driven outflows the pressure exceeds
average pressure in the host galaxy gas by a factor as large as a few
tens. They suggested that the result of this high ambient gas pressure is the
formation of very compact star clusters by over-pressurising Giant Molecular
Clouds (GMC). The sizes of the resulting clusters are comparable to that of
the Globular Clusters. Similarly, we suggest that over-pressuring of the gas
in the dwarf galaxy caused by the passage of the host galaxy's outflow may
initiate formation of Globular Clusters in dwarf galaxies. Recently,
\cite{Assmann2013} proposed that dwarf galaxies could have formed from the
merger of star clusters in low-mass dark matter haloes. The passage of an
outflow would provide a natural way to trigger the clustered star formation
which is the basis of those models.

It is also possible that the material ejected from the dwarf, e.g.,
gas at radii $r > r_S$, clumps up to form stars, as long as it is able to cool
rapidly enough. This may create a population of globular clusters at large
galactocentric radii whose gas originated from dwarf satellites of the host.

There is a testable prediction that this picture makes which we hope could be
checked observationally in future. Namely, starbursts in satellite galaxies
induced by an outflow from a much larger host galaxy are co-eval within the
time required for the outflow to sweep the host galaxy (e.g., tens to $\sim
100$ Million years), a time that is very short compared with the age of the
host and the satellites. Assuming that the starbursts produce observationally
significant amount of stars and/or globular clusters per dwarf this suggests
that there should be co-eval peaks in the star formation history of seemingly
unrelated dwarf galaxies. Current data are consistent with all dSphs having
had an early burst of star formation more than 10 Gyr
ago~\citep[e.g.][]{Tolstoy2009}, although the uncertainties associated with
the ages of stellar populations make it difficult to establish whether these
bursts were exactly co-eval. The range of star formation histories exhibited
by the dSphs in their subsequent
evolution~\citep[e.g.][]{deBoer2012a,deBoer2012b}, may indicate that a hybrid
model involving both AGN-driven shocks and interaction with the hot halo are
required to understand fully the Milky Way's dSph population.

\section{Discussion and Conclusions}

We have noted that the ram pressure of galactic outflows driven by either
quasars/AGNs or powerful starbursts in a host galaxy is large enough to
affect the gas in the satellite dwarf galaxies orbiting the host. We first
demonstrated the point for a simple, singular isothermal sphere model for the
host and the satellite galaxies (\S \ref{sec:toy}), and we then considered a
more realistic \cite{NFW} potential. Considering separately the gas settled
in a rotationally supported disc, and also a gaseous hot halo of the dwarf
galaxy, we calculated the fraction of gas ejected from the satellite by the
host's outflow and the fraction of gas retained (see Figs. 2 and 3 in \S
\ref{sec:real}). As an example, we found (see fig. 2) that for a Milky Way
like host, a small dwarf (with maximum circular velocity $v_{\rm circ} =15$ km
s$^{-1}$) loses practically all its gas if it is within $\sim 100$ kpc
of the centre of the host galaxy. The same occurs for distances closer than
$\sim 30$ kpc for a much more massive satellite with $v_{\rm circ} = 60$ km
s$^{-1}$. In \S \ref{sec:mw} we considered a well defined sample of dSph
galaxies of the Milky Way, using the latest available data for their orbits in
the Galactic halo and assuming NFW models for the dSphs' internal structure. We found that
many, if not most, of the dSphs in the sample could well have been affected by
the putative Galactic outflow; unfortunately, the orbital data in particular
are still not accurate enough, which leaves significant room for uncertainty
for most of dSphs.

We now compare the overall energetics of our outflow stripping the Galaxy of
most of its gaseous mass out to its virial radius with the likely feedback
energy sources: the SMBH named Sgr~A$^*$ and the stellar population of the
Galaxy. The total amount of kinetic energy in the most energetic model we
studied here -- the NFW shock -- is
\begin{equation}
E_{\rm sh} = M_{\rm sh} {V_{\rm
    sh}^2\over 2} \approx 10^{16} \msun \hbox{(km/s)}^2\;,
\label{esh}
\end{equation}
where $M_{\rm sh} \approx 10^{11}\msun$ is the total mass of the shell driven
outward to the virial radius of the Galaxy. This estimate is made for $V_{\rm
  sh} = 500$~km~s$^{-1}$.  Now, for \sgra, the black hole mass is $M_{\rm BH}
= 4.4\times 10^6\msun$ \citep{GenzelEG10}. The total kinetic energy released
in the fast outflow in \citep{King03} model is
\begin{equation}
E_{\rm BH} = {1\over 2} {v_{\rm AGN}\over c} \epsilon M_{\rm BH} c^2 \sim
2\times 10^{15} \msun \hbox{(km/s)}^2\;,
\label{ebh}
\end{equation}
where $\epsilon\approx 0.1$ is the standard radiative efficiency for disc
accretion \citep{Shakura73} and $v_{\rm AGN} \sim 0.1 c$ is the fast nuclear outflow
velocity \citep{KP03}.

We observe that $E_{\rm BH}$ is a factor of $\sim 5$ below the required
energy, $E_{\rm sh}$. However, if the outflow was somewhat focused along the
direction perpendicular the Galactic plane then the energy requirement could
be met along those directions. Only the satellites located in those directions
would then be affected by the ram pressure shock discussed here. In addition,
it is certainly possible that $\epsilon\sim 0.2$ if the SMBH is rotating
rather than not, and $v_{\rm AGN}$ could be somewhat larger. We could also
have over-estimated the mass of the gas in the shell escaping the
Galaxy. Therefore we believe that while \sgra\ fails to power the {\em most}
energetic shell of the two considered here with the default parameters, there
is still plenty of parameter space where feedback from \sgra\ could be
sufficient for a significant impact on the Galaxy's satellites.  

For stellar feedback, we consider only supernovae type II, for which the
\cite{Kroupa02} IMF yields the kinetic energy output of $\sim 5\times 10^{5}
\msun \hbox{(km/s)}^2$ per $1 \msun$ of the total stellar mass in the
population (this is derived assuming that each SNe releases $10^{51}$ erg of
kinetic power, and that all stars more massive than 8 $\msun$ yield type II
SNe. Now, the total stellar mass of the Milky Way is estimated at $M_{\rm
  tot}\sim 6\times 10^{10}\msun$ \citep{McMillan10}, which releases a total of
$3 \times 10^{16} \msun \hbox{(km/s)}^2$ in SN type II over the lifetime of
the Galaxy. If a significant fraction, e.g., $\sim 1/3$, of this energy were
released in the star burst, then this would be sufficient to power our most
energetic outflow.

Finally, recent detailed AGN feedback simulations \citep{NZ12} and analytical
models (Zubovas et al 2013, MNRAS submitted) show that AGN outflow may
actually trigger (at least accelerate) a very powerful star burst in a
gas-rich phase of galaxy formation. In that scenario both the SMBH outflow and
the starburst SN type II would pump the energy into the outflow clearing the
galaxy of gas. We therefore conclude that our model does not require an
unreasonable amount of energy.

The shock ram pressure stripping process considered here is yet another
plausible way to resolve the ``missing satellite'' problem of the Milky Way,
which we note could have operated {\em in conjunction} with the tidal
stripping and the ram pressure stripping of the satellite galaxies due to the
present day hot halo of the Galaxy. Having a dwarf galaxy harassed by the
Galactic outflow early on does not preclude its further harassment by the
present day hot halo \citep[e.g.,][]{MayerEtal06}, and in fact should increase
the efficiency of the latter process: a less massive gas disc is easier to
strip away from the dwarf by the hot halo's ram pressure.

We also pointed out (\S \ref{sec:induced}) that the pressure in the host
galaxy's outflow is sufficient to compress the gas in the discs of the dwarf
galaxies and thus trigger star formation there. Observationally this picture
could be tested by looking for a co-eval spike in the star formation histories
of the dwarfs of the Milky Way. Additionally, the epoch of Galactic outflow
may be the best time to form Globular Clusters -- not only in the main Galaxy but also
in the dwarf satellites, as the outflow provides both very high ram and external pressures (cf. Nayakshin and McLaughlin, 2013).

Finally, we note that we have not considered the possible collimating effect
on the outflow of the presence of a massive disk in the host galaxy. In this
case, the impact on the satellite population might not be isotropic, giving
rise to anisotropies in the distribution of detectable satellites around the
host, or variations of satellite properties depending on their locations
relative to the outflow direction. Such structures need not necessarily be
aligned with the plane of the present-day disk, as the outflow is generated
during a major merger whose angular momentum could lead to an accretion disk
around the central SMBH which was randomly oriented relative to the
larger-scale galactic disk. This effect might explain the presence of
large-scale, flattened structures in the distribution of satellite galaxies
around both the Milky Way~\citep{Pawlowski2012} and M31~\citep{Ibata2013}
which have been claimed in the literature. Further exploration of our model is
required to determine its impact on the spatial distribution of satellites
around a host.

\section{Acknowledgments}

Theoretical astrophysics research at the University of Leicester is supported
by a STFC Rolling grant. MIW acknowledges support from the Royal Society
through a University Research Fellowship.

\label{lastpage}

\end{document}